\documentclass[twocolumn,aps,prl,showpacs,superscriptaddress,amsmath,amsfonts,amssymb]{revtex4-1}
\usepackage[latin9]{inputenc}
\usepackage{amstext}
\usepackage{graphicx}
\usepackage{esint}
\usepackage{cancel}

\makeatletter
\usepackage{dcolumn}
\usepackage{bm}
\usepackage{times}
\usepackage{color}

\makeatother

\begin{document}

\title{Defects, disorder and strong electron correlations in orbital degenerate,
doped Mott insulators}

\author{Adolfo Avella}

\affiliation{Dipartimento di Fisica ``E.R. Caianiello'', Universit\`a degli
Studi di Salerno, I-84084 Fisciano (SA), Italy}

\affiliation{CNR-SPIN, UoS di Salerno, I-84084 Fisciano (SA), Italy}

\affiliation{Unit\`a CNISM di Salerno, Universit\`a degli Studi di Salerno,
I-84084 Fisciano (SA), Italy}

\author{Andrzej M. Ole\'{s}}

\affiliation{Max-Planck-Institut für Festkörperforschung, Heisenbergstrasse 1,
D-70569 Stuttgart, Germany}

\affiliation{Marian Smoluchowski Institute of Physics, Jagiellonian University,
prof. \L{}ojasiewicza 11, PL-30348 Kraków, Poland}

\author{Peter Horsch}

\affiliation{Max-Planck-Institut für Festkörperforschung, Heisenbergstrasse 1,
D-70569 Stuttgart, Germany}

\date{\today}
\begin{abstract}
We elucidate the effects of defect disorder and $e$-$e$ interaction on 
the spectral density of the defect states emerging in the Mott-Hubbard
gap of doped transition-metal oxides, such as Y$_{1-x}$Ca$_{x}$VO$_{3}$.
A soft gap of kinetic origin develops in the defect band and survives
defect disorder for $e$-$e$ interaction strengths comparable to the 
defect potential and hopping integral values above a doping dependent
threshold, otherwise only a pseudogap persists. These two regimes
naturally emerge in the statistical distribution of gaps among different
defect realizations, which turns out to be of Weibull type. Its shape 
parameter $k$ determines the exponent of the power-law dependence of the 
density of states at the chemical potential ($k-1$)
and hence distinguishes between the 
soft gap ($k\geq2$) and the pseudogap ($k<2$) regimes. Both $k$ 
and the effective gap scale with the hopping integral and the $e$-$e$ 
interaction in a wide doping range. The motion of doped holes is 
confined by the closest defect potential and the overall spin-orbital 
structure. Such a generic behavior leads to complex non-hydrogen-like 
defect states that tend to preserve the underlying $C$-type spin and 
$G$-type orbital order and can be detected and analyzed via 
scanning tunneling microscopy.
\end{abstract}

\pacs{75.25.Dk, 68.35.Dv, 71.10.Fd, 71.55.-i}

\maketitle
Defects in semiconductors and insulators determine their transport
properties and are responsible for their usefulness for electronics.
The hopping between defect states depends on their relative energy and 
is largely a function of disorder. In case of small hopping amplitudes,
the long-range $e$-$e$ interaction becomes extremely relevant as it 
modifies substantially the energy of defect states and their occupations.
In a seminal work \cite{Efr75,Efr76}, it was shown that a soft gap 
develops in the density of states (DOS), 
$N(\omega)\propto|\omega|^{\kappa}$ with exponent $\kappa=d-1$ for 
system dimension $d=2,3$, in the classical Coulomb glass model:
it is known as Coulomb gap \cite{Pol92}.
Further theoretical \cite{Epp97,Iof04,Efr11} and experimental 
\cite{But00} studies confirmed the remarkable success of the strong 
coupling approach for defects.

\begin{figure}[t!]
\includegraphics[width=8.0cm]{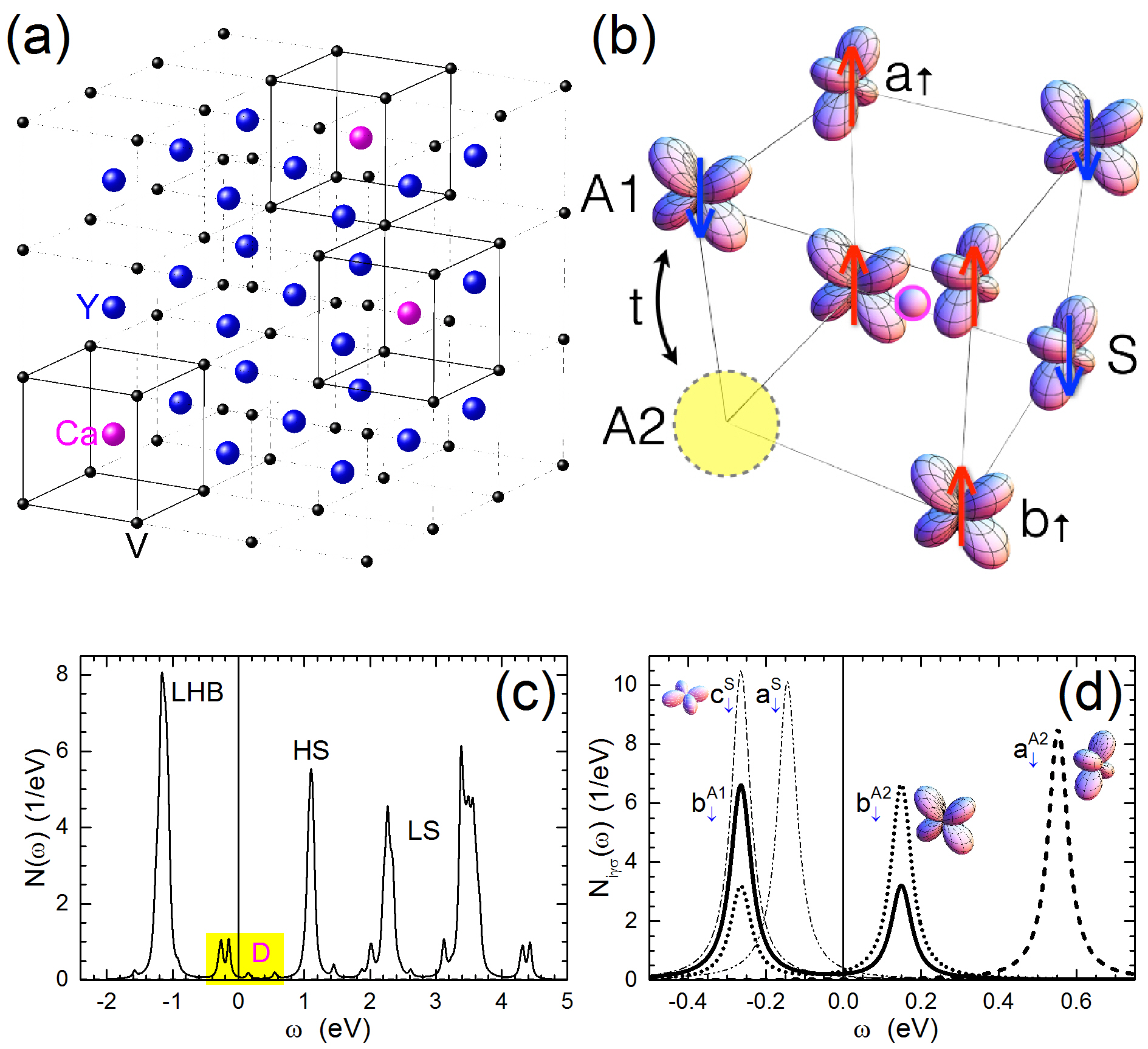} 
\protect\caption{(color online) 
(a) Y$_{1-x}$Ca$_{x}$VO$_{3}$ lattice with a random
distribution of Ca defects. (b) A Ca defect in the center of a cube
made of 8 V ions. The related hole (yellow circle) is confined to
move (hopping $t$) along a vertical bond: the \emph{active} 
$\langle$A1,A2$\rangle$ bond. The occupied $a/b$ orbitals and spin 
states obey $C$-AF spin and $G$-AO order \cite{Fuj08} on 
\textit{spectator} (S) sites.
(c) The LHB and the high-spin/low-spin
(HS/LS) states of the UHB for a periodic arrangement of defects and
$x=2\%$. The defect states D (yellow rectangle) are located within
the MH gap for $V_{{\rm D}}=1.0$ eV and $t=0.2$ eV. (d) The zoom
of the defect states D uncovers the contributions of the \emph{active}
bond and \emph{spectator} sites (heavy and thin lines)
and the formation of the kinetic gap.\label{fig:1}}
\end{figure}

We consider defects in a quite different class of compounds:
Mott insulators exhibiting a Mott-Hubbard (MH) gap due to short-range
$e$-$e$ interactions \cite{Ima98} that separates the lower Hubbard
band (LHB) from the upper Hubbard band (UHB) \cite{Mei93}. Defects
in Mott insulators feature many fascinating behaviors 
\cite{Brz15,Tan05,Kim05,Sen10,Yan11,Pav12,Jes15}
and are usually thought to lead to only two alternatives: either the
MH gap collapses or the defect states inside the gap undergo an Anderson
transition, as proposed by Mott \cite{Mot89} for La$_{1-x}$Sr$_{x}$VO$_{3}$
and for the high-$T_{c}$ cuprates. However, why the insulator-to-metal
transition occurs in vanadates at much higher doping than in cuprates,
although in both systems the MH bands do not disappear with metallization
\cite{Abb05,Fuj08}, is still not understood. Then, instead from the
Anderson-Hubbard model that features only short-range Hubbard-like
interactions and one orbital flavor \cite{Bel94,Shi09,Kot97,Son08,Saw10},
we start from an extended Hubbard model with long-range $e$-$e$ 
interactions, which allows us to study the effect of the self-consistent 
screening of defect potentials, and 3 orbital flavors. 
It provides a platform for describing the spin-orbital correlations
of the perovskite vanadates, such as Y$_{1-x}$Ca$_{x}$VO$_{3}$,
with active $\{yz,zx\}$ orbitals at V$^{3+}(xy)^{1}(yz/zx)^{1}$
ions, and coexisting $C$-type antiferromagnetic ($C$-AF) spin and
$G$-type alternating orbital ($G$-AO) order \cite{Fuj10}, 
see Figs.~\ref{fig:1}(a) and (b).

The motion of a doped hole is bound to the charged Ca defect {[}Fig.~\ref{fig:1}(b){]}
and is further controlled by the underlying spin-orbital structure:
it forms a localized spin-orbital polaron \cite{Woh09,Hor11}. Figure~\ref{fig:1}(c)
displays the associated defect states in the MH gap in the case of
a periodic arrangement of defects or, equivalently, of a short-range
defect potential \cite{Ave13}, and it also reveals the multiplets
in the UHB. Due to the $CG$ spin-orbital order, holes tend to form
dimer states on specific $c$-bonds, the \emph{active} bonds, which
results in the formation of a kinetic gap, see Fig.~\ref{fig:1}(d).
Our main goal is to understand whether this kinetic gap survives the
potential fluctuations of random defects with long-range Coulomb potentials
and which role the screening due to the $t_{2g}$ electrons plays.

Crucial to our analysis are the electron-defect ($V_{im}^{{\rm D}}$)
and the $e$-$e$ ($V_{ij}$) interactions, both screened by the background
dielectric constant $\epsilon_{c}$ due to core electrons (no $t_{2g}$
electrons), 
\begin{equation}
V_{im}^{{\rm D}}=v(R_{im}),\hskip0.3cmV_{ij}=\eta v(r_{ij}),\hskip0.3cm
v(r)=\frac{e^{2}}{\epsilon_{c}\,r},
\label{pots}
\end{equation}
where $R_{im}$ and $r_{ij}$ stand for the electronic distances between
the V ion at site $i$ and the Ca defect at site $m$ and between two V 
ions at sites $i$ and $j$, respectively. The typical binding energy of 
a hole is $V_{{\rm D}}=V^{{\rm D}}(d)\approx1$ eV \cite{Fuj08}, 
where $d$ is the distance between the defect and its closest V ions 
and $\epsilon_{c}\simeq5$. A hole would propagate along the $c$ axis
at $V_{{\rm D}}=0$ \cite{Ish05}, similar to an $e_{g}$ hole in
Y$_{2-x}$Ca$_{x}$BaNiO$_{6}$ \cite{Dag96}.

The Hamiltonian of the doped Y$_{1-x}$Ca$_{x}$VO$_{3}$ reads as
\begin{eqnarray}
{\cal H}_{t2g} & = & \sum_{im}V_{im}^{{\rm D}}n_{i}
+\sum_{i\neq j}V_{ij}n_{i}n_{j}+{\cal H}_{{\rm CF}}
+{\cal H}_{{\rm JT}}\nonumber \\
& - & \sum_{\left\langle ij\right\rangle \sigma \alpha}t_{ij}^{\alpha}(d_{i\sigma\alpha}^{\dagger}d_{j\sigma\alpha}^ {}+{\rm H.c.})
+{\cal H}_{{\rm loc}}(U,J_{H}),\label{H3band}
\end{eqnarray}
where $n_{i}=\sum_{\sigma\alpha}n_{i\sigma\alpha}$ and $n_{i\sigma\alpha}=d_{i\sigma\alpha}^{\dagger}d_{i\sigma\alpha}^ {}$,
with orbital flavor $\alpha\in\{a,b,c\}$ standing for $a\equiv yz$,
$b\equiv zx$, $c\equiv xy$. The 1$^{st}$ two terms in Eq.~(\ref{H3band})
basically resemble the Coulomb glass model \cite{Efr75,Efr76} with
site energies determined by the (random) positions of defects. The
$e$-$e$ interaction $V_{ij}$ plays a major role in determining
the occupation of these states as for $\eta=1$ the combined defect-hole
potential is dipolar \cite{notemo}, while for $\eta=0$ it is monopolar.
$V_{ij}$ is also responsible for the additional screening involving
the transitions between the Hubbard bands and the defect states. Further
terms in the 1$^{st}$ line, 
${\cal H}_{{\rm CF}}\!=\!-\Delta_{c}\sum_{i\sigma}n_{i\sigma c}$
and ${\cal H}_{{\rm JT}}$, denote the crystal-field and Jahn-Teller
terms for the $t_{2g}$ electrons \cite{Ave13}. A new dimension of
the defect problem arises from the 2$^{nd}$ line that includes the
nearest-neighbor hopping (the symmetry of $t_{2g}$ orbitals 
implies that $t_{ij}^{\alpha}$ is equal to $t$
and different from $0$ only for a bond $\langle ij\rangle$ direction
different from $\alpha$ \cite{Kha01,Aha03,Dag08}), and the local 
Hubbard physics of the triply degenerate $t_{2g}$ electrons,
${\cal H}_{{\rm loc}}(U,J_{H})$ \cite{Dag11}. The local Coulomb
interactions include intraorbital Hubbard $U$ and Hund's exchange
$J_{H}$ expressed in the SU(2) invariant form \cite{Ole83}. They
are responsible for the multiplets in the UHB for $d$-$d$ charge
excitations {[}Fig.~\ref{fig:1}(c){]}.

\begin{figure}[t!]
\includegraphics[width=7.6cm]{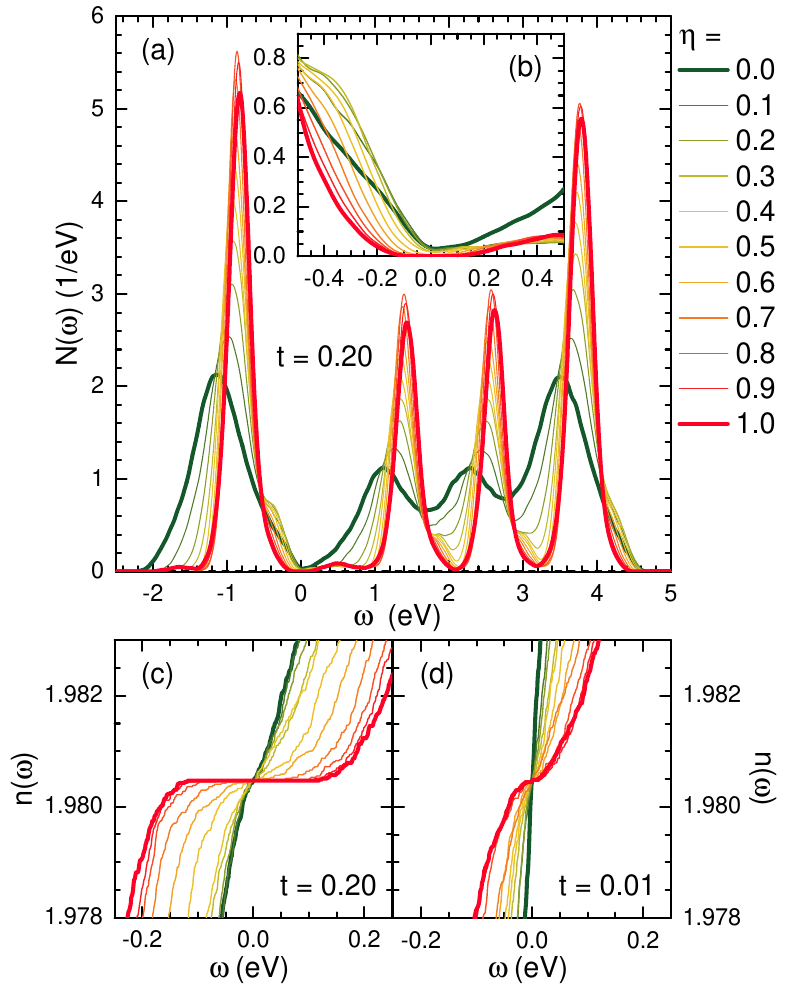} 
\protect\caption{(color online) 
(a) Density of states $N(\omega)$ averaged over $M=100$
defect realizations for doping concentration $x=2\%$, $t=0.2$ eV,
and for $\eta\in[0,1]$. A Gaussian smearing of $0.03$ eV has been
used. Inset (b) shows a zoom of $N(\omega)$ in (a) close to the Fermi
energy ($\omega=0$). A zoom of the averaged integrated DOS $n(\omega)$
close to Fermi energy is shown for: (c) $t=0.2$ eV, and (d) $t=0.01$
eV.\label{fig:2}}
\end{figure}

We solve the Hamiltonian~(\ref{H3band}) self-consistently employing
the unrestricted Hartree-Fock (uHF) approximation \cite{Miz96}. There
are two main advantages of the uHF approach we like to emphasize:
(i) uHF reproduces the Hubbard bands and the multiplet splitting not
only for undoped systems \cite{Miz96}, but also in presence of defects
\cite{Hor11} and orbital polarization and SU(2) rotation \cite{Ave13};
(ii) the spatial distribution and the occupation of each defect state
depends on all other occupied states in presence of disorder and 
long-range interactions~(\ref{pots}).
As a matter of fact, uHF solves this central and complex optimization 
problem in the most efficient way. The derivation of the uHF equations 
is standard; more details can be found, for instance, in 
Refs.~\cite{Hor11,Ave13}. We present results obtained for a cluster of
$N_{a}=8\times8\times8$ V ions with periodic boundary 
conditions, after averaging over $M=100$ statistically different 
Ca defect realizations. 
We use the standard parameters for YVO$_{3}$, i.e., $U=4.0$ eV, 
$J_{H}=0.6$ eV, $\Delta_{c}=0.1$ eV \cite{Ave13}.

The $2_{\rm spin}\times3_{\rm orbital}\times N_{a}$ uHF eigenvalues 
$\epsilon_{s,l}$ obtained for a given defect realization $s$ yield 
the averaged DOS per V ion, 
\begin{equation}
N\left(\omega\right)=\frac{1}{M}\sum_{s=1}^{M}\left[\frac{1}{N_{a}}\sum_{l=1}^{6N_{a}}\delta(\omega+\mu_{s}-\epsilon_{s,l})\right].
\label{Nw_av}
\end{equation}
The Fermi energy $\mu_{s}$ not only separates the occupied from the
unoccupied states in each defect realization $s$, but as well reflects,
via the energy optimization, a repulsion between such states as in
the Peierls effect \cite{Hor98}. Therefore, the average over different
defect realizations calls for an overall alignment of the energy scales
by means of the different $\mu_{s}$.

Figure~\ref{fig:2} displays the variation of the MH multiplets for
different strengths of $e$-$e$ interaction, encoded by the parameter
$\eta$, for doping $x=2\%$ of random Ca defects (i.e., for 10 defects)
[cf. Figs.~\ref{fig:1}(c) and (d) for a periodic arrangement of 
defects]. The electronic states close to the defects are pushed
by the potential $V_{{\rm D}}$ away from the LHB into the MH gap.
However, the actual energy distribution of defect states is strongly
dependent on the screening of the $t_{2g}$ electrons via the $e$-$e$
interaction and a\emph{ }\textit{\emph{soft gap}}\emph{ }gradually
opens up in the DOS on increasing $\eta$. The inset (b) clearly shows
the non-monotonous variation of the defect states inside the MH gap
on varying the screening. On the large energy scale, two important
changes occur when $\eta$ is varied. For $\eta=0$, the defect potential
is unscreened and the interaction with further randomly distributed
defects broadens the Hubbard bands. For $\eta=1$, the screening is
instead complete: each defect forms an exciton with a doped hole and
the resulting interaction between excitons is dipolar with a tremendous
suppression of the effects of disorder and a dramatic narrowing of
the Hubbard bands.

To analyze the behavior of the soft gap in $N(\omega)$ without
suffering from the unavoidable smearing, we discuss next the averaged
integrated DOS, $n(\omega)=\int_{-\infty}^{\omega}d\omega'N(\omega')$,
in the vicinity of the Fermi energy and the related plateau {[}see
Figs.\ \ref{fig:2}(c) and (d){]}. It is worth noting the following
key features in $n(\omega)$: (i) there is an evident gap/plateau
for $t=0.2$ eV (being a typical value for cubic vanadates \cite{Kha01})
and $\eta=1$, but not for small $t=0.01$ eV, and (ii) on decreasing
the screening $\eta\rightarrow0$, the gap/plateau disappears even
for $t=0.2$ eV.

\begin{figure}[t!]
\includegraphics[width=8.4cm]{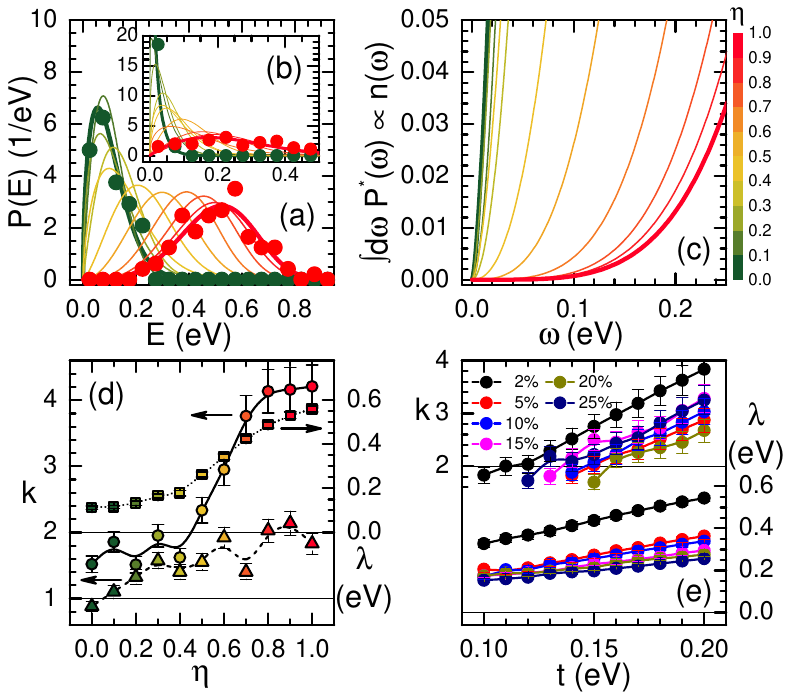} 
\protect\caption{(color online) 
(a) $P(E)$ for $t=0.2$ eV and different vaules of $\eta$ 
(colors as in Fig.~\ref{fig:2}). Lines are least-squares fits from Eq. (\ref{Wei})
and dots are numerical data for $\eta=0$ and $1$ computed from the
$M$ defect realizations; inset (b) same as (a) but for $t=0.01$ eV; 
(c) averaged integrated DOS $n(\omega)$ calculated from (\ref{Wei});
(d) $\eta$ dependence of $k$ and $\lambda$ for $t=0.2$ eV (circles
and squares, respectively) and for $t=0.01$ eV (only $k$ with triangles);
(e) $t$ dependence of $k$ and $\lambda$ for doping $x=2,5,10,15,20$ and 
$25\%$ at $\eta=1$. Lines in (d) and (e) are guides to the eye.}
\label{fig:3}
\end{figure}

In order to establish the statistical behavior of $N\left(\omega\right)$
in the limit $M\to\infty$, we use that $N\left(\omega\right)$
is proportional to the probability distribution function
$P^*\left(\omega\right)$ that a state in a generic defect realization 
has energy $\omega$ relative to its Fermi energy $\mu_{s}$.
Then, we find that a generic defect realization 
features a gap of size $E$ with a probability governed by
a Weibull probability distribution function,
\begin{equation}
P(E)=\theta\!\left(E-\zeta\right)\,\frac{k}{\lambda}\,\left(\frac{E-\zeta}{\lambda}\right)^{k-1}
\mathrm{e}^{-\left(\frac{E-\zeta}{\lambda}\right)^{k}},
\label{Wei}
\end{equation}
with shape parameter $k$, scale parameter $\lambda$ and location
parameter $\zeta$. Accordingly, if $\zeta=0,$ we have $P^*
\left(\omega\right)=\frac{k}{\lambda^{k}}\left|\omega\right|^{k-1}$
and $N\left(\omega\right)\propto\left|\omega\right|^{k-1}$ both for 
$\left|\omega\right|\ll\lambda$,
that is we have a soft gap for $k\geq2$, a pseudogap for $1<k<2$ 
and no gap for $k=1$. Instead, if $\zeta>0$, we have 
$N\left(\omega\right)=0$ for $\left|\omega\right|\leq\zeta$ and 
$N\left(\omega\right)\propto\left(\left|\omega\right|-\zeta\right)^{k-1}$
for $\zeta<\left|\omega\right|\ll\lambda$, that is we have a hard
gap. Thus, $P\left(E\right)$ results in a robust scheme to determine
the behavior of $N\left(\omega\right)$ close to the Fermi energy,
that is the presence and type of gap in the system. The numerical
data obtained for the gaps of $M$ defect realizations for
$t=0.2\;(0.01)$ eV and $\eta=0$ and 1 are compared in Figs.~\ref{fig:3}(a)
and (b) to the corresponding statistical least-squares fits to $P(E)$.
The fits are indeed excellent in all cases and give systematically $\zeta=0$.

In Fig.~\ref{fig:3}(c), we report the $n(\omega)$ curves of 
Fig.~\ref{fig:2}(c) successfully reconstructed with the help of 
$P\left(E\right)$. The plateau/gap $\Delta$ present in 
Fig.~\ref{fig:2}(c) for $\eta\geq0.5$ is due to the finiteness of 
$M$: its statistical value is $\Delta\doteq\lambda/\sqrt[k]{M}$
that vanishes for $M\to\infty$. Figures~\ref{fig:3}(d) and \ref{fig:3}(e)
summarize the dependence of $k$ and $\lambda$ on the $e$-$e$ interaction
strength $\eta$ and $t$, respectively. 
Both $k$ and $\lambda$ increase with increasing $e$-$e$ interaction $\eta$,
see Fig.~\ref{fig:3}(d). At $t=0.2$ eV, for $\eta>0.5$, we have $k>2$ and,
therefore, a soft gap. On the contrary, for $t=0.01$ eV, $k<2$ is found for
all values of $\eta$: the $e$-$e$ interaction \textit{alone} 
is not sufficient to stabilize a gap and only a pseudogap persists.
It is worth noting the almost linear increase of both $k$ and 
$\lambda$ with increasing $t$ shown at $\eta=1$ in Fig.~\ref{fig:3}(e),
which justifies calling the soft gap a kinetic gap. We also observe 
a rather slow, but monotonous, decrease of $\lambda$ on increasing the
doping $x$. The most important feature is the non-universality of
the exponent $k$ that scales with both $\eta$ and $t$, and is not 
simply given by the system dimensionality, in contrast to 
the Coulomb gap in disordered semiconductors \cite{Efr75,Efr76}. 

\begin{figure}
\includegraphics[width=8.4cm]{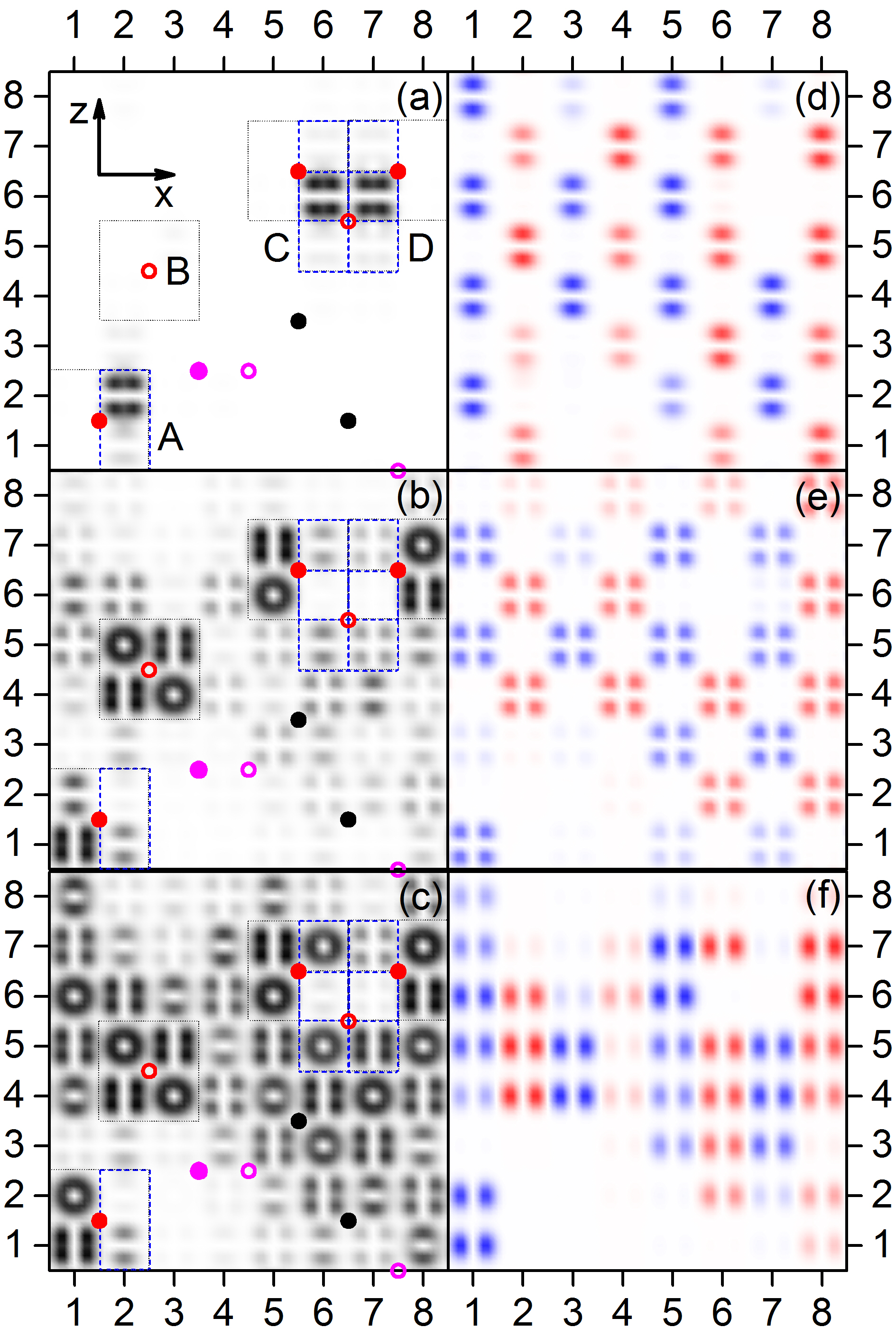}
\protect\caption{(color online) 
Integrated electron/hole density $\varrho_{\sigma\alpha}(x,y,z;V)$
in the $ac$ plane with $y=1$ {[}V ions are at $(x,1,z)$ sites{]}
for a typical defect realization at $x=2\%$, $\eta=1$, and $t=0.2$~eV. 
The defects closest to the shown plane at $y=1.5$ ($0.5$) are
marked by red dots (circles), at $y=2.5$ ($7.5$) by magenta dots
(circles) and more distant ones by black dots. Faces of V cubes hosting
a defect are indicated by thin gray dotted lines, while \emph{active}
bonds by thick blue dashed lines. Panel (a) shows the integrated unoccupied
density at $V=1.0$ eV, with defect features A, B, C, D discussed
in the text; panels (b) and (c) show the integrated occupied density
at: (b) $V=-0.7$ eV, (c) $V=-0.8$ eV. Right panels show the spin-orbital
partial densities at $V=-0.8$ eV for: (d) $a=yz$, (e) $b=xz$, and
(f) $c=xy$ orbitals. Red (blue) color for up (down) spin projections
clearly show $C$-AF order.\label{fig:4}}
\end{figure}

The kinetic gap formation is triggered by the doped holes that do not 
form symmetric, hydrogen-like, orbitals around the defects. Instead,
due to the interplay with the spin-orbital order, they form composite
spin-orbital polarons that localize in a symmetry broken form on 
\emph{active} bonds. Which of the 4 closest $c$-bonds of a defect is 
chosen depends on the interactions with all other defects. 
To detect and analyze these complex defects, we 
study in the following the scanning tunneling microscopy (STM) patterns 
\cite{Fis07,Pas00,Mul04,Law10} that correspond here to the spatially 
resolved spin-orbital ($\sigma\alpha$) DOS integrated from the Fermi 
energy to the applied voltage $V$ for a particular defect realization 
$s$, $\varrho_{\sigma\alpha}(x,y,z;V)=|\int_{0}^{V}d\omega\,
\rho_{\sigma\alpha}(x,y,z;\omega+\mu_{s})|$.

The integrated unoccupied density pattern summed over all spin-orbital
degrees of freedom, $\sum_{\sigma\alpha}\varrho_{\sigma\alpha}(x,y,z;V)$,
is shown in Fig.~\ref{fig:4}(a) for $V=1.0$ eV. In the lower left
corner, we recognize an unoccupied defect state (A) at coordinates
$(x,y,z)=(2,1,z)$ with a finite hole density at vanadium sites $z=1,2$
(on the \emph{active} bond). The asymmetry relative to its closest Ca 
defect at $(1.5,1.5,1.5)$ is evident. The degree of orbital polarization,
i.e., increased weight at $z=2$, is due to the other defects and
the Jahn-Teller potential. Fig.~\ref{fig:4}(b) shows the occupied
density for $V=-0.7$ eV. Close to the same defect at $(1.5,1.5,1.5)$,
we see two occupied $c$-bonds: one at (1,1,1\&2) with two electrons per 
site (\emph{spectator} sites), and another one at (2,1,1\&2) --- the 
\emph{active} bond (A), with a single hole fluctuating in an asymmetric
way along the bond parallel to the $c$ axis. The defect (B) has its
hole on a neighbor $y$-plane and we see only \emph{spectator} sites.
(C) and (D) mark a pair of \emph{active} bonds belonging to three
V cubes hosting three defects. More defect states appear at $V=-0.8$
eV {[}Fig.~\ref{fig:4}(c){]} that are not well separated from the
LHB. Here the complexity of the defect landscape is apparent as well
as the interaction of the doped holes with the spin-orbital background.

The landscapes in Figs.~\ref{fig:4}(d-f) represent the partly 
occupied spin-orbital densities $\varrho_{\sigma\alpha}(x,y,z;V)$ 
of \textit{\emph{defect states}} at $V=-0.8$ eV. The red/blue stripe 
structure for up (down) spins reveals that both the underlying $C$-AF 
spin order and the $G$-AO order survive the doping by charge defects, 
in contrast to what happens in high-$T_{c}$ cuprates where the spin 
order of the parent compound is destroyed \cite{Kha93,Ave91}. This 
supports the findings of the Tokura's group that $C$-AF/$G$-AO order 
is preserved in various doped vanadate systems \cite{Fuj08}.

Summarizing, we have shown that charged defects in vanadates
generate an intrinsic kinetic gap within the Mott-Hubbard
gap that survives defect disorder for strong $e$-$e$ interactions
implying a strong dielectric screening. The kinetic gap transforms into 
a soft gap with power-law dependence: 
$N(\omega)\propto\left|\omega\right|^{k-1}$. We have established that 
the exponent $k$ is non-universal and scales with both the kinetic 
scale $t$ and the $e$-$e$ interaction strength $\eta$. We suggest that 
an STM analysis can provide highly valuable microscopic information 
on the complex non-hydrogen-like states of doped holes, 
but this remains an experimental challenge.

\acknowledgments

We thank A. Rost and H. Shinaoka for insightful discussions. A.A.
acknowledges kind hospitality at Max-Planck-Institut für Festkörperforschung,
Stuttgart. A.M.O. kindly acknowledges support by Narodowe Centrum
Nauki (NCN, National Science Center) Project No. 2012/04/A/ST3/00331.

\end{document}